# Importance of both spin and orbital fluctuations in $BaFe_2(As_{1-x}P_x)_2$ : Evidence from superconducting gap anisotropy

T. Yoshida,[1,2]* S. Ideta,[1] T. Shimojima,[3] W. Malaeb,[4] K. Shinada,[3] H. Suzuki,[1] I. Nishi,[1] A. Fujimori,[1,2] K. Ishizaka,[3] S. Shin,[4] Y. Nakashima,[5] H. Anzai,[6] M. Arita,[6] A. Ino,[5] H. Namatame,[6] M. Taniguchi,[5,6] H. Kumigashira,[7] K. Ono,[7] S. Kasahara,[8,9] T. Shibauchi,[9] T. Terashima,[8] Y. Matsuda,[9] M. Nakajima,[1] S. Uchida,[1,2] Y. Tomioka,[2,10] T. Ito,[2,10] K. Kihou,[2,10] C. H. Lee,[2,10] A. Iyo,[2,10] H. Eisaki,[2,10] H. Ikeda,[2,9] R. Arita,[2,3] T. Saito,[2,11] S. Onari,[2,12] and H. Kontani[2,11]

[1]Department of Physics, University of Tokyo, Bunkyo-ku, Tokyo 113-0033, Japan.

[2]JST, Transformative Research-Project on Iron Pnictides (TRIP), Chiyoda, Tokyo 102-0075, Japan.

[3]Department of Applied Physics, University of Tokyo, Tokyo 113-8656, Japan.

[4]Institute of Solid State Physics, University of Tokyo, Kashiwa 277-8581, Japan.

[5]Graduate School of Science, Hiroshima University, Higashi-Hiroshima 739-8526, Japan.

[6]Hiroshima Synchrotron Center, Hiroshima University, Higashi-Hiroshima 739-0046, Japan.

[7]KEK, Photon Factory, Tsukuba, Ibaraki 305-0801, Japan.

[8]Research Center for Low Temperature and Materials Sciences, Kyoto University, Kyoto 606-8502, Japan.

[9]Department of Physics, Kyoto University, Kyoto 606-8502, Japan.

[10]National Institute of Advanced Industrial Science and Technology, Tsukuba 305-8568, Japan.

[11]Department of Physics, Nagoya University, Furo-cho, Nagoya 464-8602, Japan

[12]Department of Applied Physics, Nagoya University, Furo-cho, Nagoya 464-8602, Japan

*Correspondence to: yoshida@wyvern.phys.s.u-tokyo.ac.jp

In the iron pnictide superconductors, two distinct unconventional mechanisms of superconductivity have been put forth: One is mediated by spin fluctuations leading to the $s_\pm$ state with sign change of superconducting gap between the hole and electron bands, and the other is orbital fluctuations which favor the $s_{++}$ state without sign reversal. Here we report direct observation of peculiar momentum-dependent anisotropy in the superconducting gap from angle-resolved photoemission spectroscopy (ARPES) in $BaFe_2(As_{1-x}P_x)_2$ ($T_c$=30 K). The large anisotropy found only in the electron Fermi surface (FS) and the nearly isotropic gap on the entire hole FSs are together consistent with modified $s_\pm$ gap with nodal loops, which can be theoretically reproduced by considering both spin and orbital fluctuations whose competition generates the gap modulation. This indicates that these two fluctuations are nearly equally important to the high-$T_c$ superconductivity in this system.

According to theories of spin-fluctuation-mediated superconductivity, various SC gap structures, such as nodeless $s_\pm$-wave, nodal $s$-wave, nodeless $s_{++}$, and nodal $d$-wave, are realized depending on the FS geometry[1-3]. Also, "horizontal" line nodes on the hole FS with $3z^2$-$r^2$ orbital character[4,5] or line nodes of closed loops on one of the electron pockets[6] have been predicted when the three-dimensionality of the FSs is taken into account. In addition to the spin fluctuations, the orbital fluctuations have also been discussed as a possible candidate for the pairing mechanism[7-9]. Observation of orbital-polarized states in the parent compound $BaFe_2As_2$ (ref. 10) and the strong softening of the shear modulus $C_{66}$ (refs. 11,12) are indicative of orbital ordering or fluctuations and, thus, stimulate theoretical studies of the orbital-fluctuation-mediated superconductivity[7-9]. So far, there have been experimental results supporting the spin-fluctuation mechanism[13] or the orbital-fluctuation mechanism[14], but the same $s$-based $A_{1g}$ symmetry of these states allows a possible new state involving both fluctuations.[7,8]

As in the case of $KFe_2As_2$ (refs. 15,16), most of nodal SC compounds in iron pnictides show low $T_c$ < 10K. However, isovalent-doping system $BaFe_2(As_{1-x}P_x)_2$ (ref. 17) shows a clear signature of line nodes with a relatively high $T_c$ (~ 30K).[18,19] Therefore, experimental identification of line nodes in $BaFe_2(As_{1-x}P_x)_2$ would give crucial information about the pairing mechanism as mentioned above. In an angle-resolved thermal conductivity measurement[20] in the vortex state, four-fold symmetry as a function of magnetic field direction has been observed, suggestive of line nodes on the electron pocket[6,21]. As for the gap on the hole FSs, laser ARPES studies by Shimojima *et al.*[22,23] have revealed nearly isotropic and FS-independent superconducting gaps around the Z point, indicating the importance of orbital fluctuations in the pairing mechanism[7-9]. In a recent study using synchrotron radiation, on the other hand, Zhang *et al.*[24] have observed a horizontal line node[4,5] on the outer hole Fermi surface around the Z point. Because no in-plane anisotropy was found in their result as well as in Shimojima *et al*'s result, the fourfold symmetry of the thermal conductivity[20] remains to be explained. In order to clarify the origin of the anisotropic superconducting state, we have performed a systematic ARPES study of $BaFe_2(As_{1-x}P_x)_2$ covering the electron and hole FSs and various $k_z$'s, and found that line nodes or at least strong gap anisotropy exists on the inner electron FS.

Energy gaps of the hole FSs for $BaFe_2(As_{1-x}P_x)_2$ ($x$=0.30, $T_c$=30 K) around the Z point[23,24] are presented in Fig. 1. Details of the sample growth, characterization,[25] and the measurement condition of ARPES are described in Supplementary S1. We observed at least two hole dispersions, implying that two out of the three bands are nearly degenerated as in the previous studies.[23,24,26] The spectra at $T$= 15K show a dip at $E_F$ for both bands, indicating a clear gap opening. The temperature dependence of the energy distribution curves (EDCs) at the Fermi momentum $k_F$ [panel b] shows a clear gap opening even slightly above $T_c$, unlike the previous result.[24]

We have also investigated the gap anisotropy [panels c-f]. In panel d, we show an example of the EDC at $k_F$ below and above $T_c$. The symmetrized EDC below $T_c$ have been divided by those above $T_c$ and we obtain peak structure at ~ 6 meV [bottom panel]. Since this structure evolves at low temperature as a shoulder in the EDCs [upper panel], we attributed this peak to a SC peak. Similar shoulders are observed at various $k_F$ points on the outer hole FS [panel e] and their energies are nearly isotropic [upper panel of Fig. 1f], consistent with the laser ARPES result.[22] Since the Fermi velocity $v_F$ of this band is also isotropic [lower panel of Fig. 1f], the four-folded oscillation in the angle-resolved thermal conductivity measurements[20] is hardly explained by the isotropic character of the outer hole band.

Next, we present ARPES spectra of the hole FSs with various photon energies and hence for various $k_z$'s as shown in Fig. 2. The photon energies hv~ 23 eV and ~35 eV correspond to the $k_z$ of the $\Gamma$ and Z points, respectively [Fig. 2a]. While only two dispersions are observed near the Z point as mentioned above, three hole dispersions are clearly resolved near the $\Gamma$ point [Figs. 2b and 2c]. For the inner and middle FSs [Figs. 2e and 2f], the gap energies are ~ 5-8 meV and are nearly independent of photon energy, i.e., independent of $k_z$. For the outer and middle bands [Figs. 2d and 2e], the EDCs around the Z point (~ 35 eV) show a hump (at ~ 12-18 meV) with a shoulder (at ~ 7 meV). While the shoulder signifies a superconducting gap as in Fig.1, the humps for the outer and middle FSs indicate a pseudogap.[27] The SC gap is nearly independent of $h\nu$ [panel g] and, hence, the hole FSs do not have a horizontal line node at any $k_z$.

In order to investigate the possible existence of line nodes on the electron FSs, a photon energy of $h\nu$ =40 eV with a circularly polarized light was used as shown in Fig. 3, because with this polarization, signals from the inner and outer electron FSs can be distinguished due to matrix element effects as shown in panel a. In the inner electron FS [Fig. 3b], while the EDC at the Fermi surface angle $\theta_{FS}$= 80° shows a clear shift (~ 5 meV) from $E_F$, the shift of the EDC at $\theta_{FS}$ = 170° is smaller (< ~ 2meV), indicating a strong anisotropy of the superconducting gap. In Figs. 3e and 3f, the SC gap magnitude is estimated from the crossing point between the EDCs below and above $T_c$ [Figs. 3c and 3d]. These plots indicate that the inner FS has a gap minimum at the edge of the FS ($\theta_{FS}$ ~0 or 180°). We have also confirmed this strongly anisotropic SC gap for the inner electron FS with several experimental conditions. (See Supplementary Information, Figs. S2 and S3.) Although the minimum gap value in the inner FS appears to be finite, it should be remembered that ARPES has finite $k_z$ resolution (inverse of the photoelectron mean-free-path ~0.5 nm). If the line nodes are distributed in a three dimensional shape such as a loop-like node,[6,22,28] $k_F$'s with finite gaps inevitably contribute to EDCs and the gap would look finite.

Here, we illustrate a possible momentum dependence of the SC gap function from the present ARPES study in Fig.4a. The outer hole FS near the Z point does not show a signature of vertical nor horizontal line nodes. In contrast to the isotropic gap on the hole FSs, the gap minimum has been observed on the inner electron FS at the longer edge of the ellipsoid-shaped FS cross-section (Fig. 3 and Fig. 4a). The observed momentum dependence of the SC gap magnitude is difficult to be understood from the spin-fluctuation-mediated mechanism alone. The comparable sizes of the SC gaps of the three hole FSs, that is, for FSs of different orbital character indicate that the inter-orbital scattering is required for the Cooper pair formation[7-9]. Also, the disappearance of the *xy* hole FS is required in the spin-fluctuation-mediated mechanism to achieve a strongly anisotropic gap or line nodes in the electron FSs,[1,28] while we have clearly observed three hole bands including the *xy* hole FS around the *Γ* point (Fig. 2).

In order to explain the gap anisotropy for both the hole and electron FSs, we have performed a model calculation including orbital fluctuations.[7,8] As shown in Figs. 4b, 4c and 4d, we have obtained a gap anisotropy similar to the experimental results of the hole and electron FSs. In this model, when the inter-orbital scattering is moderately strong, the highly anisotropic gap of the electron FS is likely to have nodes in the crossover region between the $s_\pm$ and $s_{++}$-wave superconductivity. The realized $T_c$ will be stable during the crossover. From the gap function of the inner electron FS (Fig. 4d), a large part of the FS has an order parameter of opposite sign to that of the hole FSs (Fig.4 c). Thus, the present observation of the gap anisotropy in the electron FS indicates the realization of nearly $s_\pm$-wave SC state.

The present result of the SC gap on the hole FSs for the $x$=0.30 sample described above have been qualitatively reproduced for a slightly overdoped $x$=0.38 sample as shown in Supplementary Fig. S4, and therefore can be considered as robust features. As for the origin of the horizontal node-like feature observed by Zhang *et al.*,[24] systematic studies on samples with controlled disorder and off-stoichiometry would be required.

As possible origins of the inter-orbital interaction, electron-phonon interaction[7] and vertex correction of the Coulomb interaction[29] have been discussed so far. Irrespective of the origin of orbital fluctuations, the strongly anisotropic SC gap and the absence of a horizontal node indicate that not only spin fluctuations but also orbital fluctuations should be taken into account in the superconductivity of iron pnictide. The interplay between the spin and orbital fluctuations may depend on the doping level and on material families, which should be clarified in future systematic studies of the SC gap in a various kind of the iron pnictide superconductors.

## Acknowledgments

We are grateful to K. Kuroki and D. L. Feng for enlightening discussions. This work was supported by a Grant-in-Aid for Scientific Research on Innovative Area "Materials Design through Computics: Complex Correlation and Non-Equilibrium Dynamics" from MEXT, the Japan-China-Korea A3 Foresight Program and a Grant-in-Aid for Young Scientist (B) (22740221) from the Japan Society for the Promotion of Science. ARPES experiments were carried out at HiSOR (Proposal No. 10-B-27 and 11-B-1) and Photon Factory (Proposal No. 2009S2-005, No. 2012S2-001, and No. 2012G075).

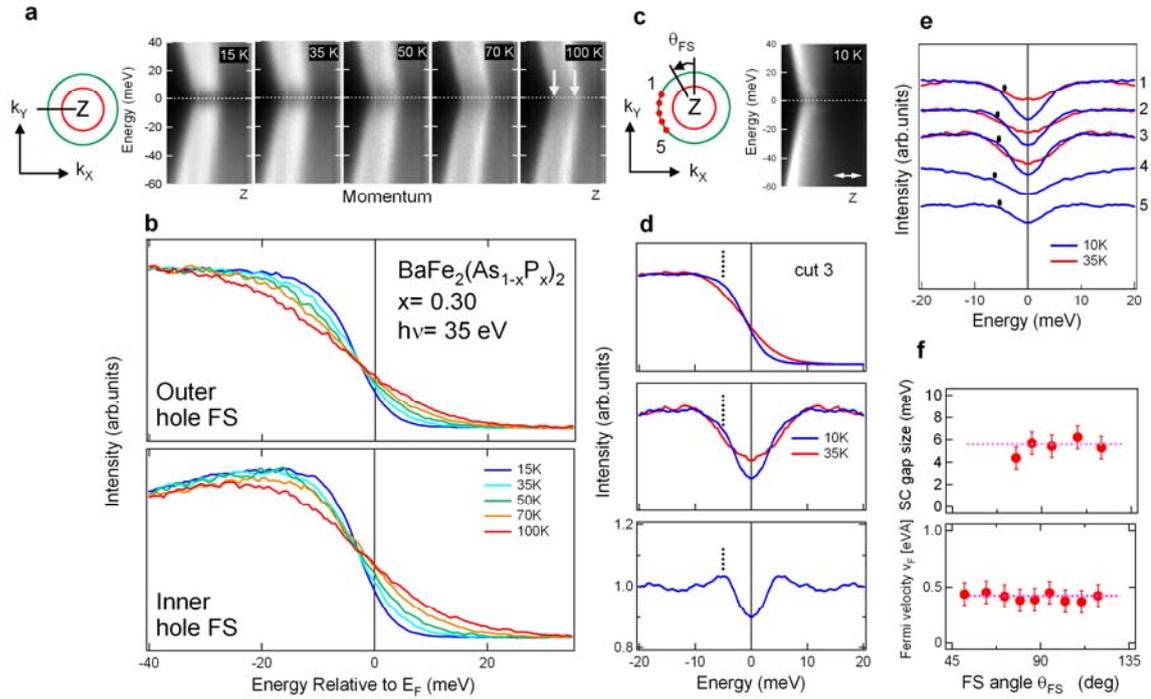

**Figure 1. Superconducting gap observed on the hole FSs in BaFe$_2$(As$_{1-x}$P$_x$)$_2$ ($x$=0.30, $T_c$=30 K) around the Z point (hν=35eV). a**, Temperature dependence of the symmetrized ARPES spectra in $k_X$ direction with a circularly polarized light. Fermi momentum $k_F$ s are indicated by arrows in the data at $T$=100K. **b**, EDCs at $k_F$ for the outer and inner hole FSs. Gaps open not only for the data below $T_c$ but also for the higher temperatures, indicating pseudogap behavior. **c**, Symmetrized ARPES spectra in $k_X$ direction (cut 3) with a linearly polarized light. The intensity of the outer hole band is enhanced due to matrix element effect. **d**, Temperature dependence of the EDC at $k_F$ above and below $T_c$ for cut 3 in panel **c**. The EDCs in the upper panel are symmetrized in the middle panel. The Spectrum in the bottom panel was obtained by dividing the spectrum below $T_c$ by that above $T_c$ in the middle panel. **e**, Symmetrized EDCs corresponding to the $k_F$ points in panel **c**. Vertical bars indicate shoulders in the low temperature spectra, indicating a superconducting gap. **f**, Sizes of the gaps determined in panel **e** are plotted as a function of Fermi surface angle $\theta_{FS}$.

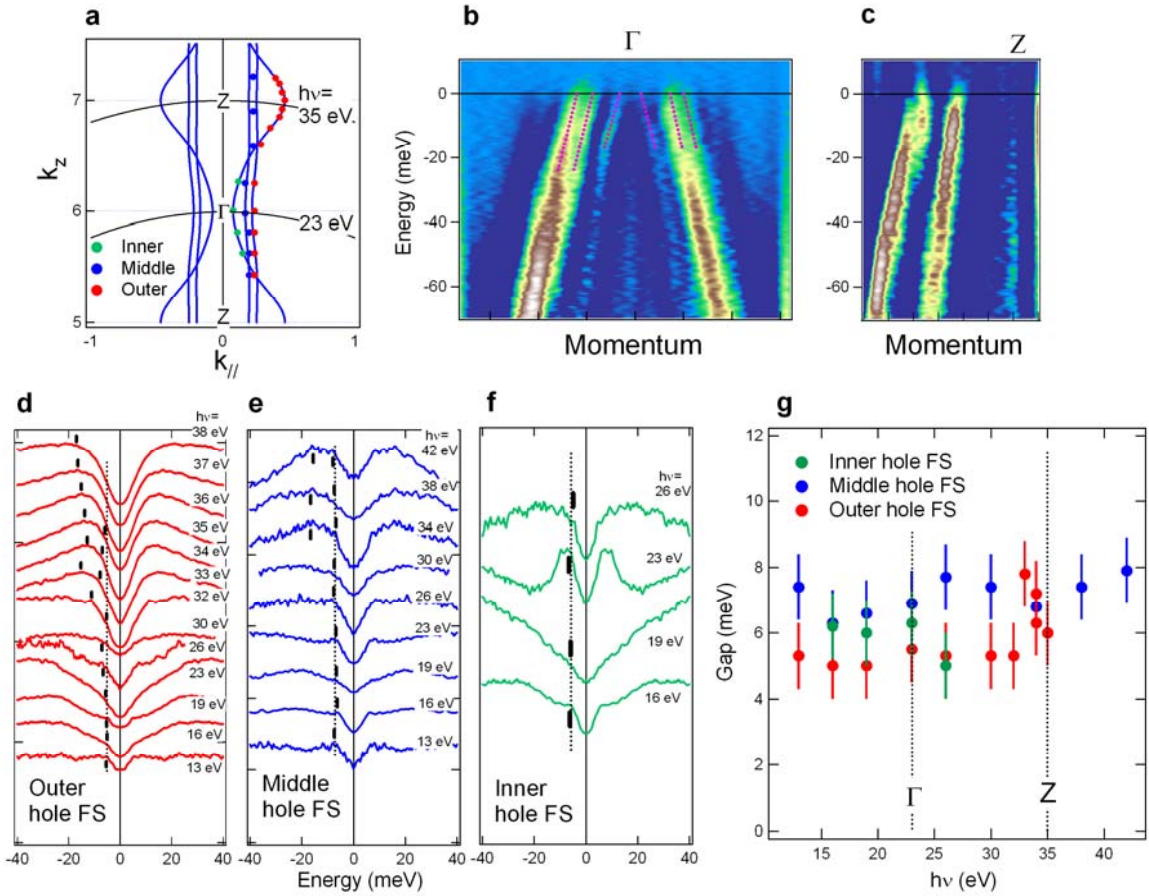

**Figure 2. Superconducting gap observed on the hole FSs with various $k_z$ in BaFe$_2$(As$_{1-x}$P$_x$)$_2$ ($x$=0.30). a**, Correspondence between incident photon energy, $k_z$ and the three-dimensional hole FSs (blue curves). **b, c**, Second derivative plots of the ARPES spectra around the $\Gamma$ and $Z$ points. Around the $\Gamma$ point, three QP dispersions are clearly resolved. **d-f**, Symmetrized EDCs at $k_F$ points in the $\Gamma$-$X$ direction taken below $T_c$. Vertical bars indicate the gap energy. **g**, Gaps estimated in panels **d-f** are plotted as a function of $h\nu$. One cannot see appreciable $k_z$ and FS-dependence.

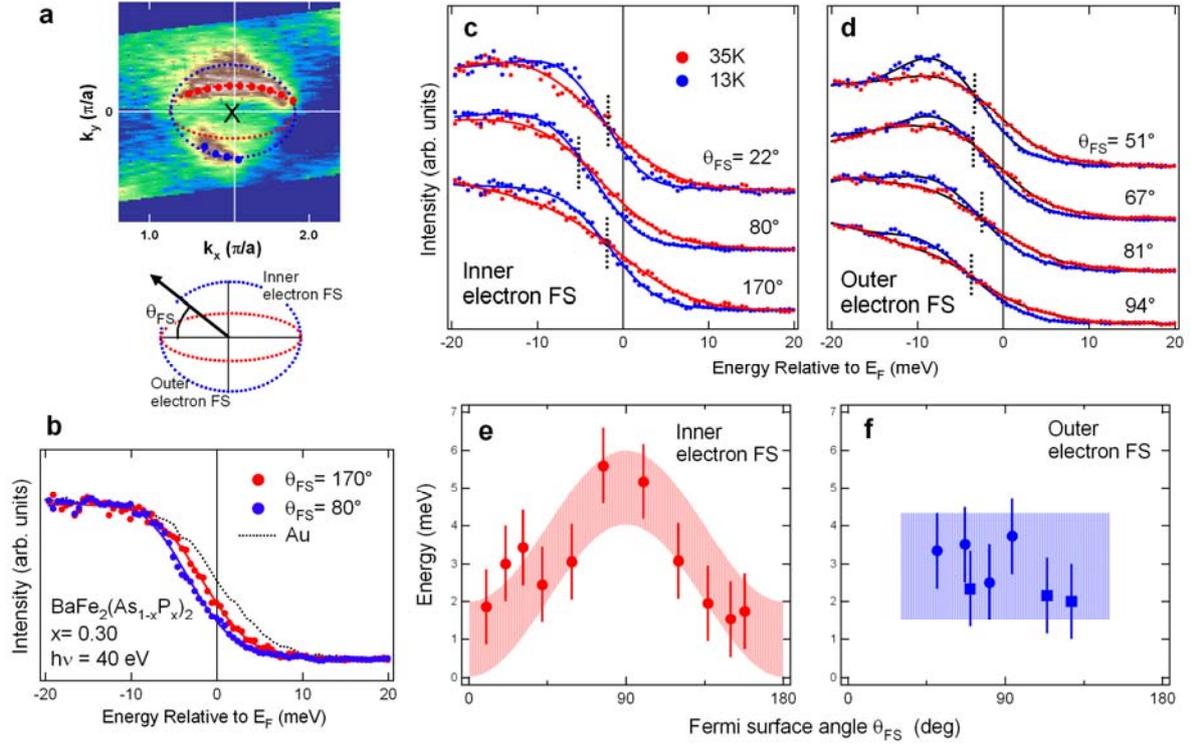

**Figure 3. Superconducting gap anisotropy observed on the electron FSs around the X point in BaFe$_2$(As$_{1-x}$P$_x$)$_2$ (*x*=0.30, *T*$_c$=30K) taken at *hν* =40 eV. a**, Fermi surface mapping using a circularly polarized light. The Fermi angle is defined so that the direction from *X* to *Γ* is *θ*$_{FS}$=0. **b**, EDCs at *k*$_F$ taken below *T*$_c$ (*T*= 13K) and compared with gold spectra. **c**, **d**, EDCs at *k*$_F$ taken below (*T*= 13K) and above (*T*= 35 K) *T*$_c$ for the inner and outer FSs. Vertical bars indicate the crossing energy between the spectra below and above *T*$_c$. **e**, **f**, Energy of the crossing point for the inner and outer FSs are plotted as a function of Fermi surface angle *θ*$_{FS}$. While the SC gap of the inner FS is highly anisotropic, a clear anisotropy is not identified in the outer FS. We could not obtain the signal of the outer band near the edge of the FS because its intensity is weak and merges with the inner band.

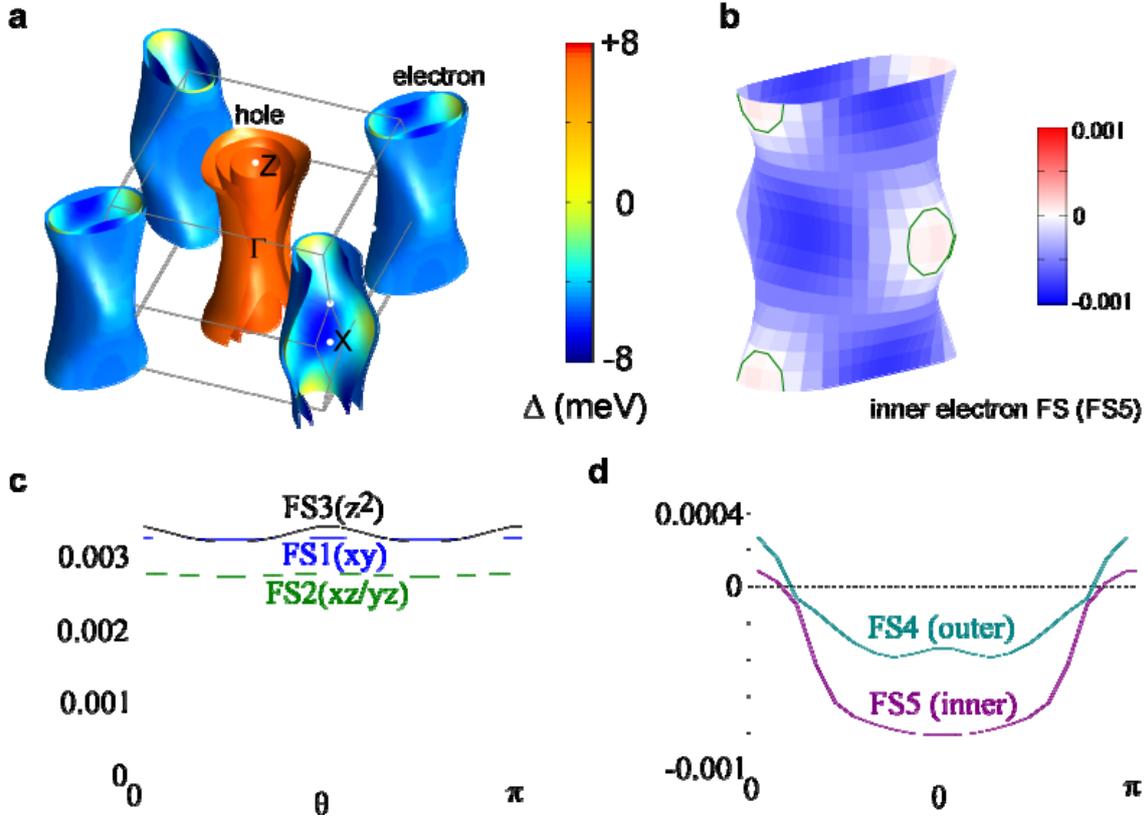

**Figure 4. SC gap $\Delta$ for all the FSs of $BaFe_2(As_{1-x}P_x)_2$. a,** Momentum dependence of the $\Delta$ for $x= 0.30$ deduced from the present ARPES result and a model calculation as below. Color scale represents the magnitude and the sign of the SC gap. Loop-like nodes or gap minima appear near the rim of the inner electron FSs. **b,** Theoretical calculation of the anisotropy of $\Delta$ for inner electron FS with orbital fluctuations.[7,8] The parameters used in the calculation, which are defined in ref. 8, are $g=0.203$ eV, $U=1.114$ eV, $\alpha_c=0.98$, $\alpha_s=0.93$, and $n_{imp}=3\%$, where g, $U$, $\alpha_{c(s)}$, $n_{imp}$ are coupling constant of a quadrupole interaction, intraorbital Coulomb interaction, charge (spin) Stoner factor, and density of impurities, respectively. **c,** Calculated SC gap function for the hole FSs around the Z point. This result indicates nearly FS-independent gap size for the hole FSs. **d,** Calculated SC gap function for the electron FSs around the X point. Both the inner and outer electron FSs have loop-like nodes. Particularly, the nodes appear near the edge of the inner electron FS, which is similar to the present ARPES result in panel **a**.

# Supplementary information

# Importance of both spin and orbital fluctuations in BaFe$_2$(As$_{1-x}$P$_x$)$_2$ : Evidence from superconducting gap anisotropy

T. Yoshida,[1,2]* S. Ideta,[1] T. Shimojima,[3] W. Malaeb,[4] K. Shinada,[3] H. Suzuki,[1] I. Nishi,[1] A. Fujimori,[1,2] K. Ishizaka,[3] S. Shin,[4] Y. Nakashima,[5] H. Anzai,[6] M. Arita,[6] A. Ino,[5] H. Namatame,[6] M. Taniguchi,[5,6] H. Kumigashira,[7] K. Ono,[7] S. Kasahara,[8,9] T. Shibauchi,[9] T. Terashima,[8] Y. Matsuda,[9] M. Nakajima,[1] S. Uchida,[1,2] Y. Tomioka,[2,10] T. Ito,[2,10] K. Kihou,[2,10] C. H. Lee,[2,10] A. Iyo,[2,10] H. Eisaki,[2,10] H. Ikeda,[2,9] R. Arita,[2,3] T. Saito,[2,11] S. Onari,[2,12] and H. Kontani[2,11]

[1]Department of Physics, University of Tokyo, Bunkyo-ku, Tokyo 113-0033, Japan.

[2]JST, Transformative Research-Project on Iron Pnictides (TRIP), Chiyoda, Tokyo 102-0075, Japan.

[3]Department of Applied Physics, University of Tokyo, Tokyo 113-8656, Japan.

[4]Institute of Solid State Physics, University of Tokyo, Kashiwa 277-8581, Japan.

[5]Graduate School of Science, Hiroshima University, Higashi-Hiroshima 739-8526, Japan.

[6]Hiroshima Synchrotron Center, Hiroshima University, Higashi-Hiroshima 739-0046, Japan.

[7]KEK, Photon Factory, Tsukuba, Ibaraki 305-0801, Japan.

[8]Research Center for Low Temperature and Materials Sciences, Kyoto University, Kyoto 606-8502, Japan.

[9]Department of Physics, Kyoto University, Kyoto 606-8502, Japan.

[10]National Institute of Advanced Industrial Science and Technology, Tsukuba 305-8568, Japan.

[11]Department of Physics, Nagoya University, Furo-cho, Nagoya 464-8602, Japan

[12]Department of Applied Physics, Nagoya University, Furo-cho, Nagoya 464-8602, Japan

1. **Materials and Methods**

Single crystals of $BaFe_2(As_{1-x}P_x)_2$ with $x=0.30$ ($T_c=30$ K) were grown by a self-flux method, adopting $Ba_2As_3$ and $Ba_2P_3$, FeAs, and FeP as starting materials (1). The precursors were mixed in the molar ratio of $Ba_2As_3$ : $Ba_2P_3$ : FeAs : FeP= 2.85 : 0.15 : 0.95 : 0.05, and then sealed in a quartz tube. All the processes were carried out in a glove box filled with dry $N_2$ gas. The tube was heated at 1150 °C for 10 h, and slowly cooled down to 900 °C at a cooling rate 1 °C/h, followed by decanting the flux. The composition of the grown single crystals was confirmed by the energy dispersive X-ray analysis.

Figure S1(a) shows temperature dependence of the in-plane resistivity. It exhibits a sharp superconducting transition at 30 K with the transition width less than 0.5 K. The magnetic susceptibility as a function of temperature was measured in a 10 Oe magnetic field parallel to the crystallographic $c$ direction [Fig. S1(b)]. A sharp drop in the zero-field-cooled (ZFC) susceptibility, indicating the onset of superconductivity, appears at 30 K with the transition width less than 0.5K (10%-90% value), consistent with the temperature where zero-resistivity appears [Fig. S1(a)] and ensuring good sample quality.

Angle-resolved photoemission (ARPES) experiments were carried out at BL 9A of Hiroshima Synchrotron Radiation Center (HiSOR) and BL 28A of Photon Factory (PF). At HiSOR BL 9A, a Scienta SES-R4000 analyzer and a circularly-polarized light were used with the total energy resolution of ~7-8 meV. At PF BL-28A, a Scienta SES-2002 analyzer and a circularly- and linearly- polarized light were used with the total energy resolution of ~8-10 meV. The crystals were cleaved in situ at $T$=8-13 K in an ultra-high vacuum of ~5 x$10^{-11}$ Torr. Calibration of the Fermi level ($E_F$) of the samples was achieved by referring to that of gold. In-plane ($k_x$, $k_y$) and out-of-plane electron momenta ($k_z$) are expressed in units of $\pi/a$ and $2\pi/c$, respectively, where $a$ = 3.92 Å and $c$= 12.8 Å are the in-plane and the out-of-plane lattice constants. Here, the $x$ and $y$ axes point towards nearest neighbor Fe atoms. Correspondence between $k_z$ and $h\nu$ is determined in our previous ARPES studies (2, 3).

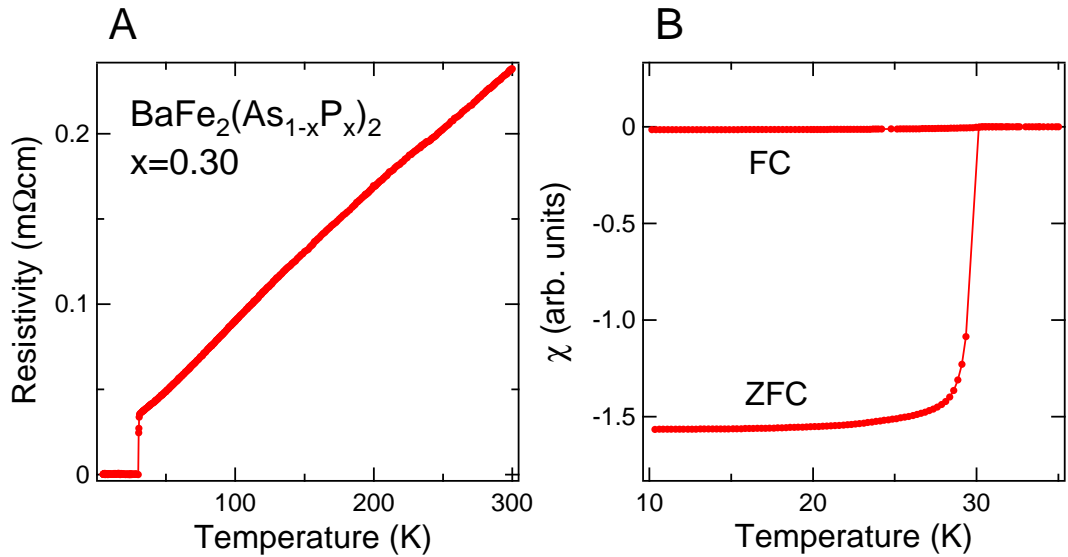

**Fig. S1**: (A) Temperature dependence of the in-plane resistivity of $BaFe_2(As_{1-x}P_x)_2$ ($x=0.30$). (B) Temperature dependence of magnetic susceptibility in a 10 Oe magnetic field along the $c$ axis.

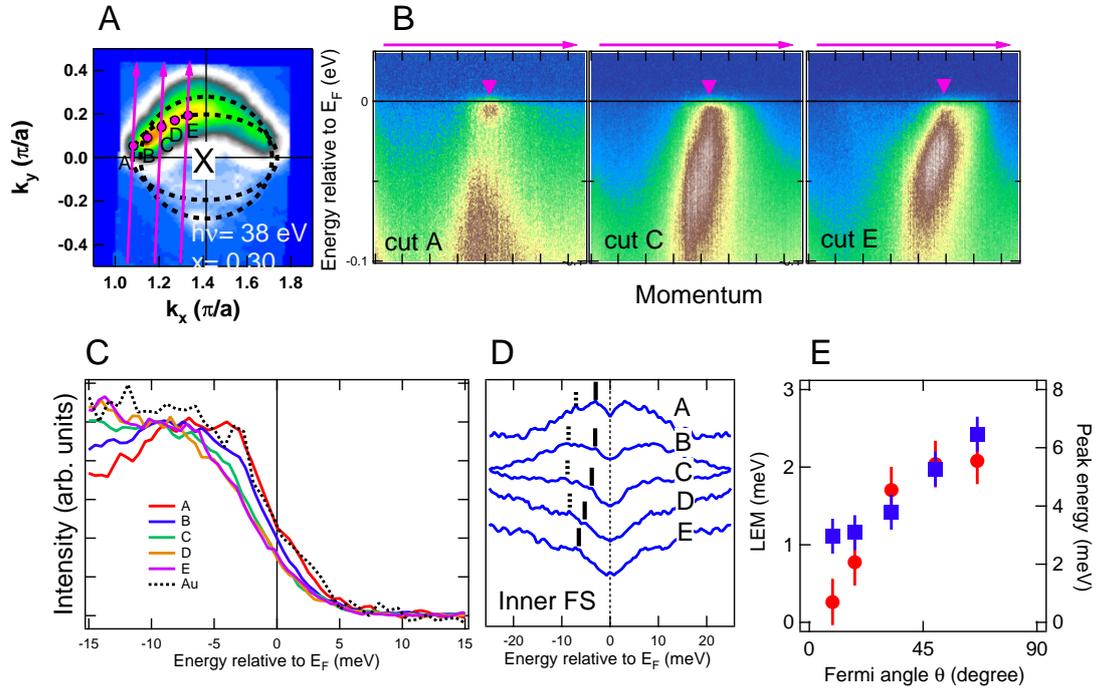

**Fig. S2**: Superconducting gap observed for the electron FSs around the $X$ point in BaFe$_2$(As$_{1-x}$P$_x$)$_2$ ($x$=0.30). (A) Fermi surface mapping for the electron FSs taken at $h\nu$=38eV. (B) ARPES intensity plot near $E_F$. Arrows indicate the $k_F$ points for the inner F S. (C) EDCs at $k_F$ points for the inner FS taken at T=10 K. (D) Symmetrized EDCs in panel (C). (E) Energy of the LEM and the peak in panels (C) and (D).

## 2. Anisotropic gap in the electron Fermi surface

Here, we show that the strongly anisotropic superconducting gap observed in the inner FS presented in the main text was also confirmed for another photon energy and another composition. Figure S2(A) shows a FS mapping for the electron bands of BaFe$_2$(As$_{1-x}$P$_x$)$_2$ ($x$=0.30) around the $X$ point taken with $h\nu$ =38eV with a circularly polarized light. The total energy resolution was ~7meV in this measurement. As shown in Fig.S2 (B), the intensity of spectral weight from the outer electron FS is suppressed due to matrix-element effects of the circularly polarized beam. EDCs at $k_F$ taken at $T=8$ K ($< T_c$) for the inner electron FS are shown in panel (C) and are compared with gold spectra. The leading edge mid-point (LEM) of point A is almost the same as that of the gold spectra, indicating a small gap. To remove the effect of temperature broadening of the Fermi-Dirac function, the EDCs at $k_F$ are symmetrized with respect to $E_F$ in panel (D). Thus obtained symmetrized EDCs clearly designate the momentum dependence of the superconducting gap. While the cut near the $X$ point [point E] indicates a clear gap opening of 6-7 meV at the peak position, the peak indicated by vertical bars is shifted towards $E_F$ in going from near-$X$ point to the edge of the FS and eventually the gap becomes less than 4 meV at the edge of the FS [point A]. As shown in the plot of LEM and the peak energy in Fig. S2(E), this result clearly indicates anisotropic superconducting gap in the inner electron pocket.

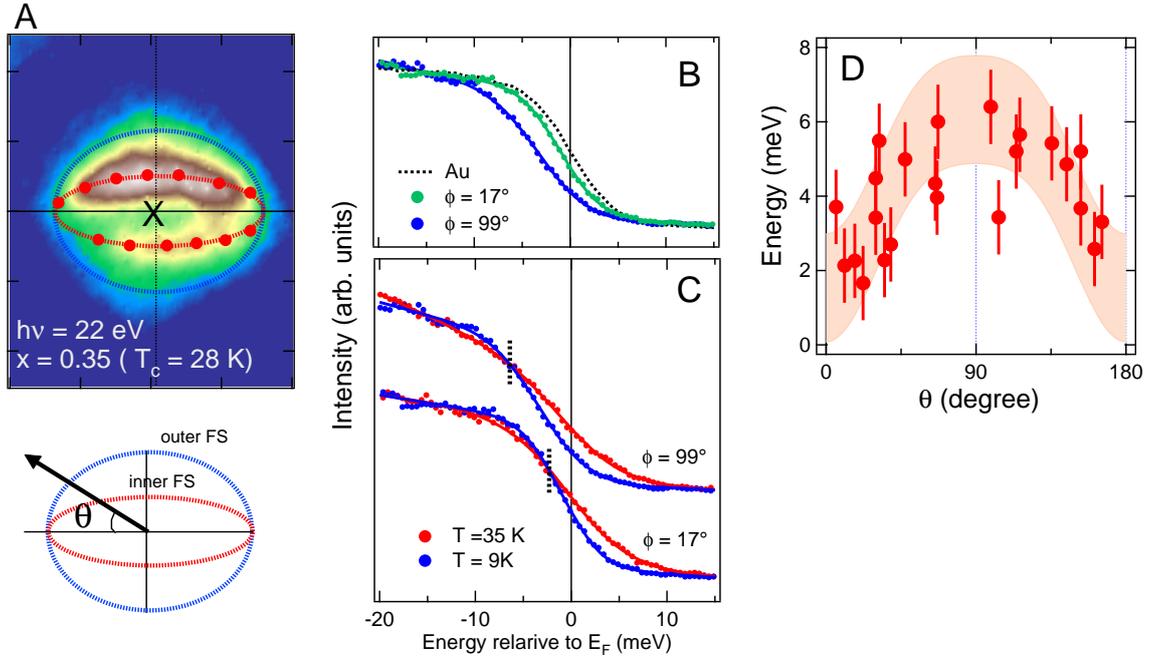

**Fig. S3**: Anisotropic superconducting gap observed on the inner electron FSs around the $X$ point in $BaFe_2(As_{1-x}P_x)_2$ ($x=0.35$, $T_c=28$ K). (A) Fermi surface mapping taken at $hv = 22$eV and schematic FSs and the definition of Fermi angle $\theta$ where the direction from $X$ to $\Gamma$ is defined as 0. (B) EDCs at $k_F$ taken below $T_c$ ($T=9$K). (C) EDCs at $k_F$ taken below ($T=9$K) and above ($T=35$ K) $T_c$. Vertical bars indicate the crossing energy between the spectra below and above $T_c$. (D) Energy of the crossing point are plotted as a function of Fermi angle.

We also show the data of superconducting gap for the inner electron FS taken at a photon energy of $h\nu$ =22eV as shown in Fig. S3. Similar to the data in Fig. S2, the intensity of the outer band is suppressed due to the matrix element effect. EDCs at $k_F$ near the *X* point and the rim of the FS are compared with the Fermi edge of Au in Fig. S2 (B). In Fig. S2(C), EDCs below (*T*= 9K) and above (*T*= 35K) $T_c$, while the EDCs at the Fermi surface angle $\theta$= 99 degree on the inner FS shows a clear shift (~ 8 meV) in going from above to below $T_c$, and the shift at $\theta$ = 17 degree is very small ( < ~ 3meV), indicating strong anisotropy of the superconducting gap. In Fig. S2 (D), we have plotted the energy shift of the crossing point between the EDCs below and above $T_c$ [Fig. S2 (C)]. This plot indicates that the inner FS has the gap minimum at the edge of the FS ($\theta$ ~ 0 or 180 degree).

## 3. Superconducting gap in the hole Fermi surface at the Z point with *x* = 0.38 sample

We have performed an ARPES experiment on the overdoped sample (*x* = 0.38, $T_c$ = 26 K) in order to check the possible existence of the horizontal node reported by Y. Zhang *et al*. (ref. 4). In Fig. S4, we show the ARPES result of the superconducting gap on the hole FSs taken at $h\nu$ = 35 eV corresponding to the Z point. Energy-momentum plot corresponding to the cut shown in Fig. S4(D) shows the outer, middle, and inner hole bands separately [Fig. S4(A), S4(B)]. In Fig. S4(C), Symmetrized EDCs at the $k_F$ taken below Tc are divided by those taken above *Tc* (T= 30 K) and these spectra clearly show superconducting peaks. Particularly, the existence of the clear superconducting peak in the outer hole band is the same with the present result of the *x* = 0.30 sample and is different from the result in Zhang et al., (ref 4). If the horizontal node exists in the overdoped samples, it could appear in the doping region larger than x=0.38.

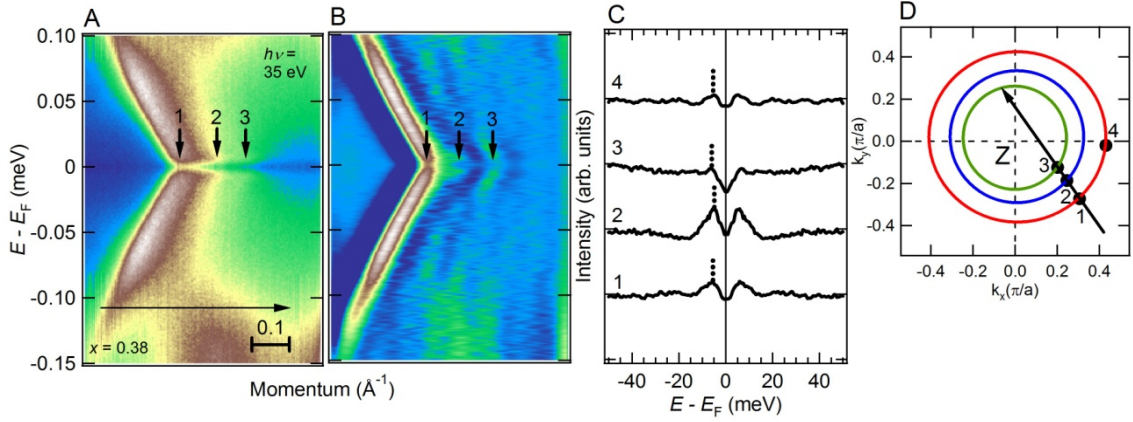

**Fig. S4:** Superconducting gap of the $x = 0.38$ sample taken at $T = 10$ K and $h\nu = 35$ eV. The $k_F$'s of the outer, middle, and inner hole bands are shown by arrows 1, 2, and 3. (B) Second-derivative energy-momentum plot of (A). (C) EDCs at $k_F$ in the superconducting state. Dotted line shows the peak position of the EDC. (D) Schematic hole FSs around the Z point. The number of 1, 2, 3, and 4 shows the $k_F$ position. A black arrow and points 1, 2, and 3 are the momentum cut and the $k_F$ points in panel (A), respectively.